\documentclass{ifacconf}
\usepackage[T1]{fontenc}
\usepackage[utf8]{inputenc}
\usepackage{graphicx}      % include this line if your document contains figures
\usepackage{natbib}        % required for bibliography
\usepackage{amsmath,amssymb,amsfonts}
\usepackage[most]{tcolorbox}
\usepackage{xcolor}
\usepackage{colortbl}  
\usepackage{array} 
\usepackage{bm}

\newtheorem{assumption}{\textbf{Assumption}}
\begin{document}
\begin{frontmatter}

\title{MPC for tracking for anesthesia dynamics}

% \title{Style for IFAC Conferences \& Symposia: Use Title Case for
%   Paper Title\thanksref{footnoteinfo}} 
% Title, preferably not more than 10 words.

\thanks[footnoteinfo]{This work was supported by the Clinical project, funded by the ANR under grant ANR-24-CE45-4255.}

\author[First]{Maxim Raymond}  
\author[Second]{Kaouther Moussa} 
\author[Third]{Mirko Fiacchini}
\author[Second]{Jimmy Lauber}

\address[First]{\textit{UPHF, CNRS, UMR 8201 - LAMIH, }
F-59313 Valenciennes, France, (e-mail: maxim.raymond@uphf.fr).}

\address[Second]{\textit{UPHF, CNRS, UMR 8201 - LAMIH,} \textit{INSA Hauts-de-France,}
F-59313 Valenciennes, France, (e-mail: kaouther.moussa@uphf.fr, jimmy.lauber@uphf.fr)}

\address[Third]{\textit{Univ. Grenoble Alpes, CNRS,} \textit{ Grenoble INP, GIPSA-lab,}
38000 Grenoble, France, (e-mail: mirko.fiacchini@gipsa-lab.fr)}

\begin{abstract}                % Abstract of 50--100 words
% In this paper, an MPC for tracking formulation is proposed for the closed-loop control of anesthesia. It seamlessly enables the optimization of 
% the steady-states pair that is not unique due to the multi input single output nature of the model. Anesthesia dynamics is a multi-time scale system 
% with two types of states caracterized, respectively, by fast and slow dynamics.
% In contrast to standard control applications, where the objective is typically to accelerate slow dynamics, 
% in anesthesia control, the output equation depends only on the fast dynamics; that one needs to accelerate during a time-window defined by the operation duration. 
% Therefore, the slow states can be treated as disturbances, and appropriate compensation terms can be introduced to account for their effect. 
% Such compensation techniques allow to reformulate the system as a nominal one (without disturbance), 
% and therefore to formulate an MPC for tracking strategy accounting for the time-varying steady states. 
% The presented framework allows to ensure recursive feasibility and asymptotic stability for the nominal system, 
% through the design of standard terminal ingredients in the MPC for tracking framework. 
% The performance of the controller is then assessed on a patient in a simulation environment.

In this paper, an MPC for tracking formulation is proposed for the control of anesthesia dynamics. It seamlessly enables the optimization of 
the steady-states pair that is not unique due to the MISO nature of the model. 
Anesthesia dynamics is a multi-time scale system with two types of states characterized, respectively, by fast and slow dynamics. 
In anesthesia control, the output equation depends only on the fast dynamics.
Therefore, the slow states can be treated as disturbances, and compensation terms can be introduced. 
Subsequently, the system can be reformulated as a nominal one allowing the design of an MPC for tracking strategy. 
The presented framework ensures recursive feasibility and asymptotic stability, 
through the design of appropriate terminal ingredients in the MPC for tracking framework. 
The controller performance is then assessed on a patient in a simulation environment.
\end{abstract}


\end{frontmatter}
%===============================================================================

\section{Introduction}

   During a surgery, the anesthesiologist is responsible for the drug scheduling of the patient.
   Post-operative comorbidities or trauma can happen due to an excess or lack of medication.
   The issue of avoiding such complications has attracted the interest of the control community.
   Anesthesia includes 3 major aspects: hypnosis which describes the depth of consciousness, analgesia which represents the lack of pain,
   and finally areflexia, the lack of movement.
   In this paper we focus on the hypnosis effect. 
   
   A well assessed indicator of the hypnosis level is the Bispectral Index (BIS). 
   This indicator is computed from the EEG signal, and ranges from a score of 100 which represents a patient fully awake, to a score
   of 0 in absence of cerebral activity.
   Different models have been proposed to represent the drug effect on the BIS using propofol alone, for example in \cite{eleveldPharmacokineticPharmacodynamicModel2018}, or by coupling it with an analgesic, for example in \cite{bouillonPharmacodynamicInteractionPropofol2004}.
   It has been proved in \cite{westDesignEvaluationClosedLoop2018} that accounting for both propofol and remifentanil to control the BIS improved the overall controller performance in comparison to a feedback
   relying solely on propofol.
   Usually, the coupled control of propofol and remifentanil is handled using a ratio as in \cite{merigoOptimizedPIDControl2019}.
   This approach may not be optimal, considering that the solution to the output equation is not unique.
   Moreover, this ratio may have been computed for a given steady-state and therefore, may not be suitable for the transient phase, which is known to be critical in anesthesia monitoring.

   For the anesthesia problem, the choice of MPC techniques is interesting due to two main reasons. Firstly, the MPC is able by design to handle constraints on
   the input and the state of the system. 
   Secondly, since the anesthesia monitoring problem may involve different objectives, their formulation in terms of a cost function may be suitable for that purpose.
   
   In the literature, different MPC frameworks have been proposed to control the depth of hypnosis, 
   for example, a robust control of anesthesia using only propofol has been presented in \cite{ionescuRobustPredictiveControl2008} using MPC.
   In \cite{ionescuNonlinearDynamicsPatients2015}, the authors proposed an MPC framework with a linear multiple input single output model obtained by means of a linearization of the output mapping.
   A nonlinear MPC has been proposed in \cite{aubouin-pairaultOnlineIdentificationPharmacodynamic2024} to regulate the BIS using propofol and remifentanil.
   To mimic the anesthesiologist behavior, in \cite{milanesiHumanImitatingControlDepth2024}, they developed an MPC controller combined with an event-based PID strategy.
   For the same purpose, in \cite{raymondMPCbasedAnesthesiologistsImitating2025}, they suggest an MPC formulation with a range cost to reduce the sensitivity of control profiles to measurement noise.
   Such methods are welcome by anesthesiologists as clinical test using MPC have been performed in \cite{pawlowskiExperimentalResultsMPC2023}.

   The considered anesthesia dynamics is a MISO system with two inputs. 
   Due to this configuration, the steady-state pair is not unique. A first issue is how to consider the complete set of admissible steady-states corresponding to a given output set-point 
   % and how to choose the right pair inside it.
   and how to optimally compute the steady-states pair during surgical operations.
   In the literature, to overcome this challenge many works rely on a fixed ratio between the two inputs. In \cite{aubouin-pairaultOnlineIdentificationPharmacodynamic2024}, although the remifentanil dosage is not regulated using a ratio, the equilibrium input computation is based on a cost that contains a soft ratio between propofol and remifentanil rates.
   As a consequence, in both cases, a unique steady-state is considered in a set-point tracking formulation, which does not represent the clinical reality.
   Moreover, such configurations may limit the flexibility in the anesthesia control process. For more vulnerable patients, anesthesiologists might want to reduce the maximum allowable drug concentrations to avoid complications, and this reduction may not be compatible
   with the same desired ratio.
   Another important aspect of anesthesia dynamics is that it is a multi-time scale system were some states have a slower response time compared to the others.
   In the case of anesthesia, the goal is to accelerate the fast states dynamics since they are those impacting the output.

   This paper aims to propose a method to design a ratio-free multiple input control of anesthesia dynamics, using propofol and remifentanil.
   To do so, a new model formulation for anesthesia is presented by considering the whole set of admissible steady-states in the tracking formulation.
   To achieve this goal, we propose a reformulation of the nonlinear output equation which, for a given fixed output set-point, results in an affine equation linking the steady-states corresponding to propofol and remifentanil drug rates.
   
   Furthermore, due to the multi-time scale dynamics related to anesthesia dynamics, 
   the slow states (concentrations in muscles and fat compartments) can be considered as a disturbance on the fast states (concentrations in blood and effect-site compartments), 
   as previously mentioned in \cite{zabiNewApproachControl2015}. Therefore, in order to cope with this problem, we design a control law that allows to compensate for the disturbance effects on the fast dynamics. 
   This compensation assumes the knowledge of the concentrations in the slow compartments, which we consider to be measured as a first step in this paper. Estimation techniques can be applied in the future, 
   similarly to what have been done in \cite{moussaDataBasedExtendedMoving2023}, \cite{aubouin-pairaultComparisonMultipleKalman2024}, to provide a more realistic solution for this problem.
   
   Leveraging the MPC for tracking framework, we propose a new anesthesia control formulation that offers more flexibility in the control design, by seamlessly integrating the admissible steady-state set into the optimization problem and 
   considering the steady-state as an additional decision variable.
   Furthermore, a supplementary cost expressing the clinical preferences allowing to guide the decision on the steady-state can be added to the objective function, without affecting the theoretical guarantees.

   This paper is organized as follows: first the model used to describe anesthesia dynamics is presented in Section~\ref{Section:Model} with the pharmacokinetics and the pharmacodynamics.
   It is followed by the new closed-loop control formulation in Section~\ref{Section:ControlLaw}. In this section the slow state effect compensation and the
   MPC design are presented. To assess the performance of the method proposed in this work, Section~\ref{Section:Results} presents the simulation setup with the numerical results.
   Lastly Section~\ref{Section:Conclusion} poses the conclusion and offers future research directions.

\section{Dynamical Model}\label{Section:Model}
   
   The physiological model used to describe anesthesia is separated into two parts,
   pharmacokinetics, which describes the drugs concentration dynamics in the different compartments of the body,
   and pharmacodynamics, which models the relation between the drug concentrations and the effect on the patient.
   Here we focus only on the hypnosis component of anesthesia monitoring.
   Therefore, the considered output is the BIS and the system inputs are the infusion rates of propofol and remifentanil.

   \subsection{Pharmacokinetics}
      Each drug pharmacokinetics is decoupled and composed of four compartments,
      that can be characterized as slow (muscles, fat)
      and fast (blood, effect-site) compartments because of the notable difference in time response.
      In the case of anesthesia the considered effect-site is the brain.
      The model parameters depend on the patient characteristics (height, weight, age, sex).
      Such expressions have been inferred from clinical data processed in different studies: \cite{eleveldPharmacokineticPharmacodynamicModel2018}, \cite{eleveldAllometricModelRemifentanil2017}, \cite{mintoInfluenceAgeGender1997}, \cite{schniderInfluenceAgePropofol1999}.
      In this paper the parameters expressions are those given by \cite{eleveldPharmacokineticPharmacodynamicModel2018} for propofol and \cite{eleveldAllometricModelRemifentanil2017} for remifentanil.
      
      In \cite{zabiNewApproachControl2015}, the authors discussed the difference in time response between the fast and slow states,
      with the muscles and fat being, respectively, 10 and 100 times slower with respect to the blood and effect-site compartments.
      In that regard they proposed to separate fast and slow dynamics. Therefore, the dynamics can be written according to the following equations:
      {\footnotesize
         \begin{align*}
            %fast
         \begin{split}
            \begin{pmatrix}
               \dot{x}_{1}(t) \\
               \dot{x}_{4}(t)
            \end{pmatrix}
            & =
            \begin{pmatrix}
               -(k_{10}+k_{12}+k_{13}) & 0 \\
               k_{e}  & -k_{e}
            \end{pmatrix}
            \begin{pmatrix}
               x_{1}(t) \\
               x_{4}(t)
            \end{pmatrix}
            +
            \begin{pmatrix}
               \frac{1}{V_{1}} \\
               0
            \end{pmatrix}
            u(t) \\
            &
            +
            \begin{pmatrix}
               k_{12} & k_{13} \\
               0 & 0
            \end{pmatrix}
            \begin{pmatrix}
               x_{2}(t) \\
               x_{3}(t)
            \end{pmatrix},
         \end{split} \\
         %slow
         \begin{split}
            \begin{pmatrix}
               \dot{x}_{2}(t) \\
               \dot{x}_{3}(t)
            \end{pmatrix}
            & =
            \begin{pmatrix}
               k_{21} & 0 \\
               k_{31} & 0 \\
            \end{pmatrix}
            \begin{pmatrix}
               x_{1}(t) \\
               x_{4}(t)
            \end{pmatrix}
            +
            \begin{pmatrix}
               -k_{21} & 0 \\
               0 & -k_{31} \\
            \end{pmatrix}
            \begin{pmatrix}
               x_{2}(t) \\
               x_{3}(t)
            \end{pmatrix},
         \end{split}
         \end{align*}
         }

      with $k_{10}=\frac{Cl_{1}}{V_{1}}$, $k_{12}=\frac{Cl_{2}}{V_{1}}$, 
      $k_{13}=\frac{Cl_{3}}{V_{1}}$, $k_{21}=\frac{Cl_{2}}{V_{2}}$, $k_{31}=\frac{Cl_{3}}{V_{3}}$, 
      where $V_{i}$ are the volumes and $Cl_{i}$ ($i=1,2,3$) the clearance rates of each compartment.
      Their computation is based on the patient's characteristics and is different for each anesthetic agent.
      The parameter $k_e$ related to the effect-site concentration has a different expression for each drug.
      The state $x \in \mathbb{R}^4$ stands for the drug concentration in each compartment, with $x_1, x_4$ corresponding to the blood and effect-site, and $x_2, x_3$ to the fat and muscles.
      % The state $x^f = (x_1, x_4)^T \in \mathbb{R}^2$ stands for the drug concentration in the blood and effect-site, and the state $x^s = (x_2, x_3)^T \in \mathbb{R}^2$ stands for the drug concentration in the fat and muscles.
      The input $u$ is the drug infusion rate.
      The units corresponding to propofol and remifentanil rates are, respectively, $[mg/s]$ and $[\mu g/s]$. 
      %For the propofol $u$ will be in $[mg/s]$ and for remifentanil $u$ will be in $[\mu g/s]$.
      In the following, we denote by $p,r$ the propofol and remifentanil pharmacokinetics state, and by $u_p, u_r$ the propofol and remifentanil pharmacokinetics input.
      We can then combine both pharmacokinetics to derive the following dynamics:
      \begin{align*}
        \dot{x}^f(t) & = A_fx^f(t) + Bu(t) + A_sx^s(t) \\
        \dot{x}^s(t) & = A_{ss}x^s(t) + A_{sf}x^f(t)
      \end{align*}
      with,
      \begin{align*}
         A_f & = \begin{pmatrix}
               A_p^f & 0_{2 \times 2} \\
               0_{2 \times 2} & A_r^f
              \end{pmatrix},
         B = \begin{pmatrix}
               B_p & 0_{2 \times 1} \\
               0_{2 \times 1} & B_r
            \end{pmatrix}, \\
         A_s & = \begin{pmatrix}
               A_p^s & 0_{2 \times 2} \\
               0_{2 \times 2} & A_r^s
            \end{pmatrix},
      A_{ss} = \begin{pmatrix}
                  A_p^{ss} & 0_{2 \times 2} \\
                  0_{2 \times 2} & A_r^{ss}
               \end{pmatrix}, \\
      A_{sf} & = \begin{pmatrix}
                  A_p^{sf} & 0_{2 \times 2} \\
                  0_{2 \times 2} & A_r^{sf}
               \end{pmatrix}, 
      % x^f(t) = \begin{pmatrix} x_p^f(t) \\ x_r^f(t) \end{pmatrix}, 
      % x^s(t) = \begin{pmatrix} x_p^s(t) \\ x_r^s(t) \end{pmatrix}
      x^f(t) = \begin{pmatrix} p_1(t) \\ p_4(t) \\ r_1(t) \\ r_4(t)\end{pmatrix}, 
      x^s(t) = \begin{pmatrix} p_2(t) \\ p_3(t) \\ r_2(t) \\ r_3(t) \end{pmatrix}
      \end{align*}
      where $u(t) = \left(u_p(t),u_r(t) \right)^T \in \mathbb{R}^2$ is the control input composed of propofol and remifentanil infusion rates and
      $x^f, x^s$ are the pharmacokinetics fast and slow states.
      %with $p, r \in \mathbb{R}^4$ respectively the propofol and remifentanil pharmacokinetics states. 
      Finally, $A_f, A_s, A_{sf}, A_{ss}, B$ are the state and input matrices for fast and slow states.
      The system is then discretized using Euler forward discretization in order to be used in the control feedback loop.

   \subsection{Pharmacodynamics}

      As stated in the introduction, the Bispectral Index (BIS) is the output quantifying the hypnosis state of the patient.
      The pharmacodynamics used in this paper will be the one from \cite{bouillonPharmacodynamicInteractionPropofol2004} as it considers the combined effect of propofol and remifentanil on the BIS.
      Moreover, they conclude that the interaction of propofol and remifentanil on the BIS is additive.
      The BIS expression is a non-linear function mapping the effect-site concentrations through a Hill function:
      {\footnotesize
      \begin{equation*}
      % y(t) = E_0 - E_{max}\left(\dfrac{\left(\dfrac{U_{p}(t)+U_{r}(t)}{U_{50}(\phi)}\right)^\gamma}{1+\left(\dfrac{U_{p}(t)+U_{r}(t)}{U_{50}(\phi)}\right)^\gamma}\right),
      y(t) = E_0 - E_{max}\left(\dfrac{\left(U_{p}(t)+U_{r}(t)\right)^\gamma}{1+\left(U_{p}(t)+U_{r}(t)\right)^\gamma}\right), 
      \end{equation*}
      }
      \begin{equation*}
         \begin{matrix}		
            U_{p}(t) = \dfrac{x^f_2(t)}{C_{e_{50}}^p}, \quad U_{r}(t) = \dfrac{x^f_4(t)}{C_{e_{50}}^r}, \\
            %\\
            %\phi = \dfrac{U_{p}(t)}{U_{p}(t) + U_{r}(t)}, \quad U_{50}(\phi) = 1 - \beta\phi + \beta\phi^2,
         \end{matrix}
      \end{equation*}
      with $C_{e_{50}}^p, C_{e_{50}}^r$ being the half-effect concentration of propofol and remifentanil, and $\gamma$ representing the slope of the Hill surface. $E_0, E_{max}$
      stand for the initial value of the BIS and drugs maximal effect. Finally, $y(t)$ represents the BIS value.
      We can reorder this equation similarly to \cite{ionescuRobustPredictiveControl2008} to obtain the following expression:
      \begin{equation}
         U_{p}(t)+U_{r}(t) = \left( \dfrac{E_0 - y(t)}{E_{max} - E_0 + y(t)} \right)^{\dfrac{1}{\gamma}}
         \label{Eq:InvertOutput}
      \end{equation}
      The commonly considered output set-point in anesthesia is 50, as it is in the middle of the targeted safe range [40, 60].
      One can notice that when considering a fixed set-point for the output $y(t)$, 
      (\ref{Eq:InvertOutput}) becomes an affine function of $U_p$ and $U_r$. This can be geometrically illustrated by the intersection of the Hill surface model and the horizontal plane at $y(t) = y_{ref} = 50$.  
      As a result we obtain the following equation:
      \begin{equation*}
         \dfrac{x^f_2}{C_{e_{50}}^p} + \dfrac{x^f_4}{C_{e_{50}}^r} = \left( \dfrac{E_0 - y_{ref}}{E_{max} - E_0 + y_{ref}} \right)^{\dfrac{1}{\gamma}}   
      \end{equation*}
      with $x^f_2$ and $x^f_4$ standing for the second and fourth entries of the fast state corresponding to the effect-site concentration of propofol and remifentanil.

\section{MPC for tracking \\ for anesthesia dynamics}\label{Section:ControlLaw}

      In this section, we first discuss a control law that compensates the slow states disturbances in Section~\ref{subSectionMPC:ControlLaw}.
      Based on this control law, we are able to formulate an MPC for tracking in Section~\ref{subSectionMPC:MPC} with theoretical guarantees discussed in Section~\ref{subSectionMPC:StabilityFeasibility}.
   
   \subsection{Control law definition with compensation}\label{subSectionMPC:ControlLaw}

      Firstly, we consider the following assumption:
      \begin{assumption}
         The fast and slow compartments drugs concentrations are considered measured without any noise.
      \end{assumption}
      As mentioned in \cite{zabiNewApproachControl2015}, slow states have a bigger time response compared to fast states. 
      Moreover, in anesthesia monitoring, unlike in many other control applications, the objective is to accelerate the fast dynamics. 
      For this reason, the slow states can be treated as a disturbance acting on the fast state dynamics.
      Accordingly, we can consider the input as a sum between a tracking term and a disturbance rejection term:
      \begin{equation*}
         u_k = v_k + m_k
      \end{equation*}
      with $v_k$ being the tracking input, and $m_k := Dx_k^s$ the disturbance rejection term aiming to cancel the slow states impact on the fast ones.
      By rewriting the state dynamics with this input formula we can find the expression of $D$:
      \begin{equation*}
         x^f(k+1) = A_fx^f(k) + Bv_k + (A_s + BD)x^s(k)
      \end{equation*}
      To cancel the effect of $x_k^s$, 
      it can be proved that, for the particular problem structure and since  $B^T B$ is invertible, then, for $D = - (B^T B)^{-1} B^T A^s$
      it follows that $(A_s + BD) = 0$.
      % it is necessary to have $(A_s + BD) = 0$ that is obtained if $B^T BD = - B^T A_s$ holds.
      % This is true for the considered structure of $B$ and $A_s$. In addition, we have that $B^T B$ is invertible,
      % and the full perturbation rejection is achieved with
      % $D = - (B^T B)^{-1} B^T A^s$.
      Then, the resulting dynamics of the fast state is: % here is the nominale sy
      \begin{equation}
         x^f(k+1) = A^f x^f_k + B v_k
         \label{Eq:NominalSys}
      \end{equation}

      Furthermore, consider the following assumption on the slow state dynamics:
      \begin{assumption}\label{Ass:slowBound}
         The slow states concentrations are considered bounded.
      \end{assumption}

      As a consequence, the slow state concentrations compensation term $Dx_k^s$ is also bounded as follows:
      \begin{equation*}
            m \in \mathcal{M}:= \{ m \in \mathbb{R}^2 : -\underline{m} \le m \le 0 \}
      \end{equation*}
      and it can be verified that, for the case under analysis, $\underline{m}$ is much smaller than $u_{max}$.

      Assumption~\ref{Ass:slowBound} is reasonable in practice since the slow states impact on fast states is usually very small, 
      although their cumulative impact over time is not negligible which is the main motivation of compensating such effects in this work.
      
      We can then define the admissible tracking input set as follows:
      \begin{equation*}
         \mathcal{V} := \mathcal{U} \ominus \mathcal{M}
      \end{equation*}
      with the operator $\ominus$ denoting the Pontryagin set difference.

   \subsection{MPC for tracking formulation}\label{subSectionMPC:MPC}

      % MPC for tracking has been introduced in \cite{limonMPCTrackingPiecewise2008} and \cite{krupaModelPredictiveControl2024} to deal with possibly varying set-point.
      % For possibly varying set-point, there is a need to recompute for each set-point the terminal ingredients, which can leads to a loss of feasibility.
      % They solved this problem by integrating the steady-states in the optimization problem as decision variables. 
      % With these new decision variables, they can express the artificial reference reached by the chosen steady-state. 
      % This artificial reference can then be used in an offset cost function to get as close as possible
      % to the desired possibly varying set-point. Thanks to that, the feasibility is not lost.
      % This technique, used in many situations, is adapted in this work to control the depth of hypnosis.

      MPC for tracking was firstly introduced by \cite{limonMPCTrackingPiecewise2008} to address the varying output set-point tracking problem within a comprehensive framework, 
      ensuring stability and recursive feasibility while circumventing the need to compute the associated ingredients online. 
      More recently, \cite{krupaModelPredictiveControl2024} presented an extensive review on the use of such techniques in various MPC-related frameworks.
      As mentioned in these works, in the case of varying output set-point, there is a need to recompute for each set-point the terminal ingredients, 
      which can lead to a loss of feasibility. The core idea of this framework is to integrate the steady-states as decision variables in the optimization problem.
      Using these decision variables, an artificial reference is defined and optimized to approximate the target reference as closely as possible by means of an offset cost.
      Thanks to this formulation, the feasibility of the framework is ensured; allowing to safely track time-varying references. 
      Such techniques have been considered in different applications. In this paper, we suggest to use the MPC for tracking framework to control 
      the depth of hypnosis in anesthesia dynamics. 

      % For the anesthesia control, the offset cost is not needed as the reference is considered constant equal to $50$.
      % Nevertheless, considering steady-states as decision variables remain interesting due to the multiple possibilities of steady-states pair.

      The MPC for tracking for anesthesia dynamics is given by:
      \begin{equation}
      \begin{aligned}
         \underset{ \bm{v},  v_a}{\text{min}} & \sum_{k=0}^{N-1}  \lVert x^f_k - x_a^f \rVert_{Q}^2 + \! \lVert v_k - v_a \rVert_{R}^2 + \! \lVert x^f_N - x_a^f \rVert_{P}^2 + \! V_d(v_a)\\
         \text{s.t.} \: & x^f_0 = x(t), \\
         & x^f_{k+1} = A^fx^f_k+Bv_k, \quad k \in \mathbb{N}_{[0, N-1]}, \\
         & (x^f_k, v_k) \in \mathcal{Z}, \qquad \qquad \:\: k \in \mathbb{N}_{[0, N-1]}, \\
         & v_a \in \mathcal{Z}_s, \\
         & (x_N, v_a) \in \mathcal{X}_a,
      \end{aligned}
      \label{Eq:MPCControl}
      \end{equation}
      with $x_a^f, v_a$ the steady-state pair, $Q \succeq 0$ and $R \succ 0$ the weighting matrices, $P \succ 0$ the terminal cost matrix, 
      $\mathcal{Z}: \{ (x, v):x \in \mathbb{R}^4, v \in \mathcal{V} \}$ the state and input constraints set, $\mathcal{Z}_s$ the set of admissible steady inputs, and $\mathcal{X}_a$ the terminal positive invariant set. Those sets are provided below. 
      $V_d(\cdot)$ represents a penalty on the steady-state input, to express clinical preferences (ratios, cost, soft constraints). 
      % Note that this $V_d$ differs from the original formulation but does not impact the theoretical proof and stability analysis.
      Note that $V_d(\cdot)$ corresponds to the offset cost used in the original formulation as will be explained subsequently.

      For the steady inputs set $\mathcal{Z}_s$, 
      % the goal here is to derive a characterization of the steady-state set with a minimal number of variables, 
      % in order to simplify the MPC related optimization problem.
      % We achieved that by rewriting $x_a^f$ in terms of $v_a$ based on the discrete equilibrium state equation:
      let's first rewrite $x_a^f$ in terms of $v_a$ based on the discrete equilibrium state equation:
      \begin{equation*}
         x_a^f = A^f x_a^f + B v_a
      \end{equation*}
      Since $A^f$ is Schur stable then, $(I_n - A^f)$ is invertible, and hence the equilibrium states can be expressed in terms of the inputs:
      % we can find the following expression of $x_a^f$:
      \begin{equation}
         x_a^f = (I-A^f)^{-1} B v_a
         \label{Eq:EquilibriumState}
      \end{equation}
      Note that the positivity of the system is ensured by imposing $v_a \ge 0$.
      The set $\mathcal{Z}_s$ corresponds to the admissible steady inputs $v_a$, for which the corresponding output set-point (BIS) is $50$. 
      The output equation at equilibrium can be rewritten using $v_a$ instead of $x_a^f$:
      \begin{equation}
         G (I-A^f)^{-1} B v_a = \left( \dfrac{E_0 - y_{ref}}{E_{max} - E_0 + y_{ref}} \right)^{\dfrac{1}{\gamma}}
         \label{Eq:OutputSteady}
      \end{equation}
      with $G = \begin{bmatrix} 0 & \dfrac{1}{C_{e50}^p} & 0 & \dfrac{1}{C_{e50}^r} \end{bmatrix}$.

\begin{figure}[ht]
         \centering
         \includegraphics[width=1\linewidth]{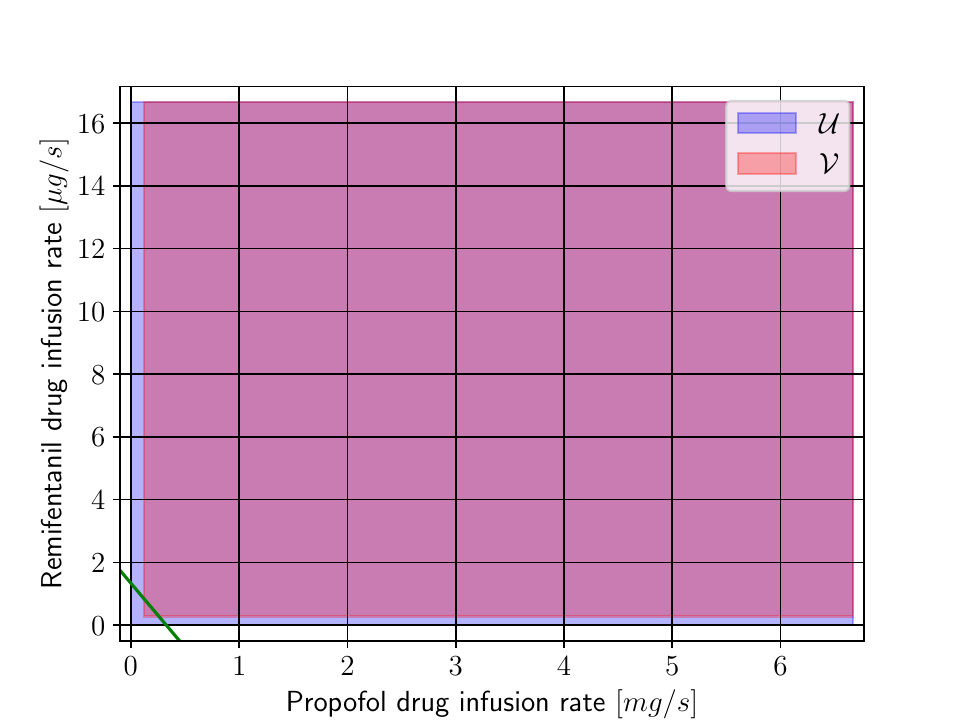}
         \vspace{-0.5cm}
         \caption{Admissible control input $u$ and MPC control input $v$ sets with the green line representing the set of $v_a$ satisfying (\ref{Eq:OutputSteady}).}
         \label{fig:EquilibriumSet}
      \end{figure}

      The set $\mathcal{Z}_s$ can therefore be defined as follows:
      \begin{equation*}
         \mathcal{Z}_s := 
         \begin{cases}
            G (I-A^f)^{-1} B v_a = \left( \dfrac{E_0 - y_{ref}}{E_{max} - E_0 + y_{ref}} \right)^{\dfrac{1}{\gamma}} \\
            u_{min} + \underline{m} + \epsilon \le v_a \le u_{max} - \epsilon
         \end{cases}
      \end{equation*}
      with $\epsilon$ positive and arbitrarily close to $0$. This $\epsilon$ is added in order to consider only strictly admissible steady-states inputs to avoid a possible
      loss of controllability, see \cite{krupaModelPredictiveControl2024} for more details.
      A representation of this set can be seen in Fig.~\ref{fig:EquilibriumSet}.

   \subsection{Stability \& Feasibility}\label{subSectionMPC:StabilityFeasibility}

      Considering the MPC setup in \cite{krupaModelPredictiveControl2024}, the asymptotic stability is proved based on the following assumption.
      \begin{assumption}
         The positive semi-definite matrix $Q$ is such that $(Q^{1/2},A^f)$ is observable. The horizon $N$ is greater than or equal to the controllability index of the system.
         The matrices $P$ positive definite and $K$ are the solution of the discrete algebraic Riccati equation.
         The set $\mathcal{X}_a$ is a positive invariant set for the state and steady inputs of the system. \label{Ass:Stability}
         The cost function $V_d(\cdot)$ is a convex, positive definite, subdifferentiable function with $V_d(0) = 0$. 
      \end{assumption}

      To find the matrices $P$ and $K$, we consider the nominal dynamics of the system described in (\ref{Eq:NominalSys}). We can define the input $v_k$ based on the error $(x_k - x_a^f)$.
      \begin{equation}
         v_k = K(x_k^f - x_a^f) + v_a
         \label{Eq:StableInput}
      \end{equation}

      Using this definition, we can rewrite the system in terms of the error:
      \begin{equation*}
         \Delta x_{k+1}^f = A^f \Delta x_k^f + B \Delta v_k
      \end{equation*}
      with $\Delta x_k^f = x_k^f - x_a^f$ and $\Delta v_k = v_k - v_a$.
      Using (\ref{Eq:StableInput}) we can find $\Delta v_k = K\Delta x_k^f$,
      and the error dynamics is, therefore, the following:

      \begin{equation*}
         \Delta x_{k+1}^f = (A^f + BK)\Delta x_k^f
      \end{equation*}

      Since $(A^f, B)$ is stabilizable then, there exist $P, K$ that satisfies the discrete algebraic Riccati equation.
      The solution $P$ will be used as the final penalty on $x^f_N$.

      The last ingredient from Assumption~\ref{Ass:Stability} is the positive invariant set.
      Based on the method proposed in \cite{limonMPCTrackingPiecewise2008},
      by substituting (\ref{Eq:StableInput}) in (\ref{Eq:NominalSys}), we have:
      \begin{equation*}
         x_{k+1}^f = (A^f + BK) x_k^f - BK x_a^f + B v_a
      \end{equation*}

      Furthermore, from (\ref{Eq:EquilibriumState}) we can rewrite the previous equation as:
      \begin{equation*}
         x_{k+1}^f = \phi x_k^f + B (I_m - \psi) v_a
      \end{equation*}
      where $\phi = (A^f + BK)$ and $\psi = K(I-A^f)^{-1}B$.

      By defining the extended state similarly to \cite{limonMPCTrackingPiecewise2008}, $w_k = \begin{pmatrix} x_k^f \\ v_a\end{pmatrix}$,
      we can rewrite the extended dynamics:
      \begin{equation*}
         w_{k+1} = A_w w_k =\begin{pmatrix}
                                 \phi & B(I_m - \psi) \\
                                 0_{m, n} & I_m 
                              \end{pmatrix} w_k
      \end{equation*}
      Given the convex polyhedron $\mathcal{W}_{\lambda}$ defined as follows:
      {\footnotesize
      \begin{equation*}
         \mathcal{W}_{\lambda} = \{ w_k = (x_k^f,v_a):(x_k,K x_k^f + (I_m - \psi) v_a) \in \mathcal{Z}, v_a \in \lambda \mathcal{V}\},
      \end{equation*}
      }
      the maximal admissible invariant set for tracking is then:
      \begin{equation*}
         \mathcal{O}^w_{\infty} = \{ w: A^i_w w_k \in W_1, \forall  i \ge 0 \}
      \end{equation*}
      Because of the unitary eigenvalues of $A_w$, we can consider the following approximation set with $\lambda$ arbitrarily close to~$1$:
      \begin{equation*}
         \mathcal{O}^w_{\infty, \lambda} = \{ w: A^i_w w_k \in W_{\lambda}, \forall  i \ge 0 \}
      \end{equation*}
      $\mathcal{X}_a$ is then defined as $\mathcal{O}^w_{\infty, \lambda}$.

      % Based on \cite{krupaModelPredictiveControl2024} the assumption needed to prove the asymptotic stability is fulfilled, and then
      % the defined MPC is asymptotically stable.

      \begin{thm}
         Consider system (\ref{Eq:NominalSys}) controlled by (\ref{Eq:MPCControl}) with Assumption~\ref{Ass:Stability} satisfied, and $x(0)$ to be in the feasibility region. 
         Then, the problem is recursively feasible, and the system is asymptotically stable and converges to $\mathcal{Z}_s$. Moreover, if $v_a^o := \arg \min \: V_d(v_a)$ is in the relative interior of $\mathcal{Z}_s$
         then, the system converges to $v_a^o$.
      \end{thm}

      \begin{pf}
         By carefully choosing the matrices $C$ and $D$ deriving from $(6)$ in \cite{krupaModelPredictiveControl2024} such that $y_a = Cx_a^f + Dv_a$ then, $V_d(\cdot)$ is equivalent to the offset cost used in the 
         original formulation of the MPC for tracking. In addition, $\epsilon$ corresponds to the $\sigma$ used in Definition~1 from \cite{krupaModelPredictiveControl2024}.
         Thus, the proof of asymptotic stability follows the same step as in \cite{limonMPCTrackingPiecewise2008}. 
         
         Considering the positive invariance of $\mathcal{X}_a$ and an optimal solution for the initial problem then, the recursive feasibility can also be proven by the same method as in \cite{limonMPCTrackingPiecewise2008}.
         $\square$
      \end{pf}

      % Concerning the feasibility, the set $\mathcal{U}$ represents the boundary of the infusion rates applied to the patient. 
      % On the other hand, the set $\mathcal{V}$ represents the boundary on the infusion rates that guarantee that $u$ is kept in $\mathcal{U}$ after adding the slow state compensation to $v$.
      % Practically speaking, there always exist $v, v_a \in \mathcal{V}$ such that we reach the desired $BIS$.
      % Consequently, if there exists an initial solution for the optimization problem of the MPC then, thanks to the positive invariant set of Assumption~\ref{Ass:Stability}
      % the problem is recursively feasible.

      Practically speaking, for the case of anesthesia, there always exist $v \in \mathcal{V}, \: v_a~\in~\mathcal{Z}_s$ such that we reach the desired $BIS$ provided that $\mathcal{Z}_s$ is not empty.
      Consequently, there always exists an initial solution for the optimization problem of the MPC.

\section{Numerical tests}\label{Section:Results}
      In order to demonstrate the effectiveness of the controller, the simulator \cite{aubouin-pairaultPASPythonAnesthesia2023} is used with a female patient
      of $56$ years old, with a height of $180\: cm$ and a weight of $92\: kg$.
      For the pharmacokinetics of propofol and remifentanil,
      the models proposed in \cite{eleveldPharmacokineticPharmacodynamicModel2018} and \cite{eleveldAllometricModelRemifentanil2017} are chosen. For the pharmacodynamics, the model of \cite{bouillonPharmacodynamicInteractionPropofol2004} is considered.
      The simulation duration is 10 minutes and the controller have a sampling time of 5s. The horizon length of the MPC $N$ is chosen to be $24$.
      The inputs constraints considered are the same as in \cite{merigoEventbasedControlTuning2020}, the propofol ranges from $0$ to $6.67$ $[mg/s]$ and the remifentanil from $0$ to $16.67$ $[\mu g/s]$.
      The complete state of the patient is considered measured.
      The matrix $Q$ and $R$ are fine-tuned in order to have a settling time smaller than 5 minutes and to minimize the undershoot. Their value are given in Table~\ref{table:Penalties_MPC} with the computed $K$ and $P$ rounded at three decimals.
      The bound $\underline{m}$ is obtained through numerical tests and its value is $(0.12, 0.27)^{T}$.
      Concerning the clinical cost $V_d(\cdot)$, it is given by:
      \begin{equation*}
         V_d(v_a) = 10 \left(v_{a_1} - \dfrac{v_{a_2}}{2}\right)^2
      \end{equation*}
      to converge to a ratio of 2, as usually done.

      \begin{table}[ht]
      \centering
      \renewcommand{\arraystretch}{1.3}

         \begin{tabular}{|cc|}
         \hline
         $Q = \left[ \begin{array}{cccc} 1 & 0 & 0 & 0 \\ 0 & 10 & 0 & 0 \\ 0 & 0 & 1 & 0 \\ 0 & 0 & 0 & 10 \end{array}\right]$, &
         $R = \left[ \begin{array}{cc} 1 & 0 \\ 0 & 1  \end{array}\right]$\\
         \hline
         \multicolumn{2}{|c|}{$K = \left[ \begin{array}{cccc} 0.671 & 1.58 & 0 & 0 \\ 0 & 0 & 0.677 & 1.267 \end{array}\right]$} \\
         \hline
         \multicolumn{2}{|c|}{$P = \left[ \begin{array}{cccc} 1.914 & 4.423 & 0 & 0 \\ 4.423 & 218.025 & 0 & 0 \\ 0 & 0 & 1.853 & 3.758 \\ 0 & 0 & 3.758 & 58.574 \end{array}\right]$}\\
         \hline
         \end{tabular}
      \caption{Controller parameters}
      \label{table:Penalties_MPC}
      \end{table}

\begin{figure}[ht]
         \centering
         \includegraphics[width=1\linewidth]{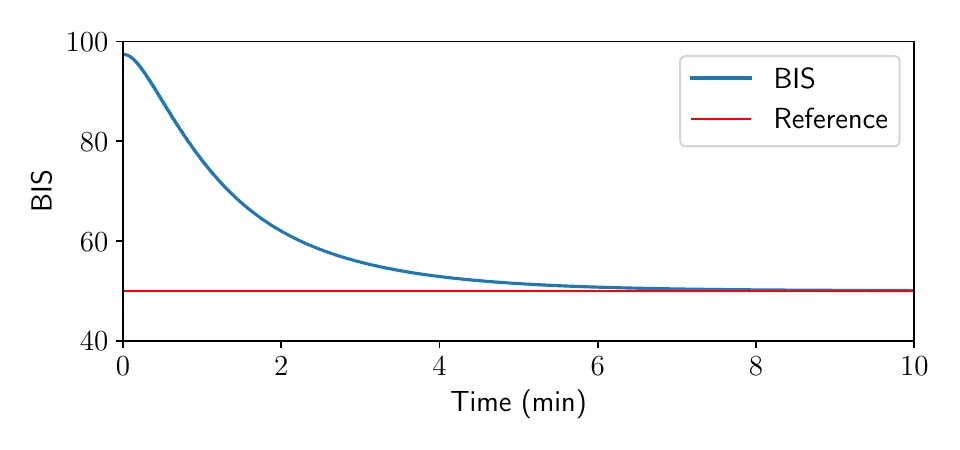}
         \vspace{-0.5cm}
         \caption{BIS evolution.}
         \label{fig:bis}
      \end{figure}

\begin{figure}[ht]
         \centering
         \includegraphics[width=1\linewidth]{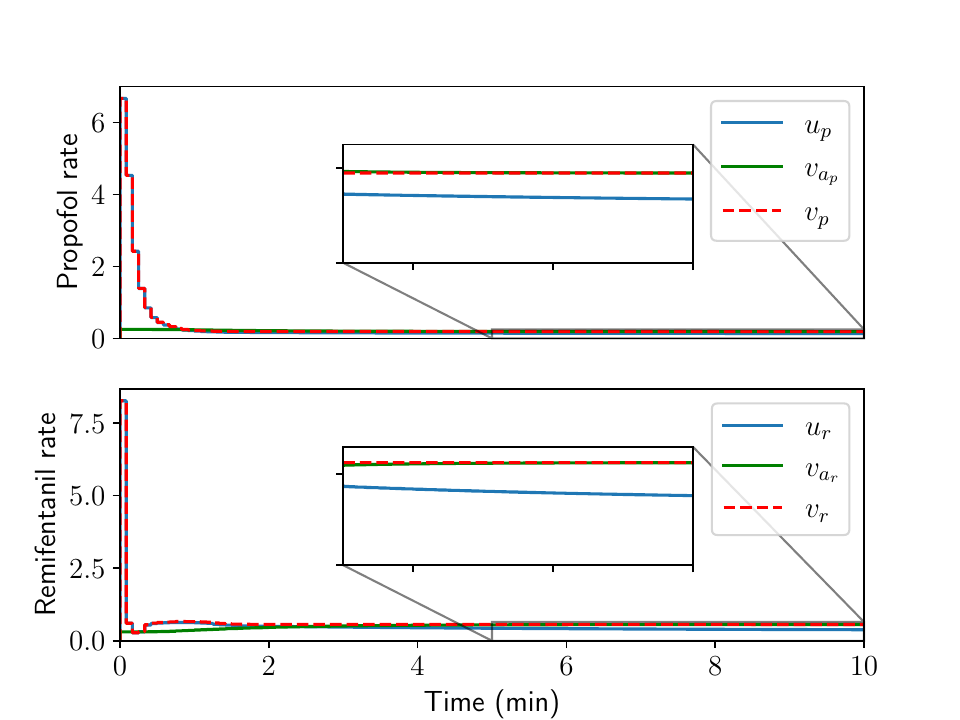}
         \vspace{-0.5cm}
         \caption{Control and steady input profiles.}
         \label{fig:input}
      \end{figure}

\begin{figure}[ht]
         \centering
         \includegraphics[width=1\linewidth]{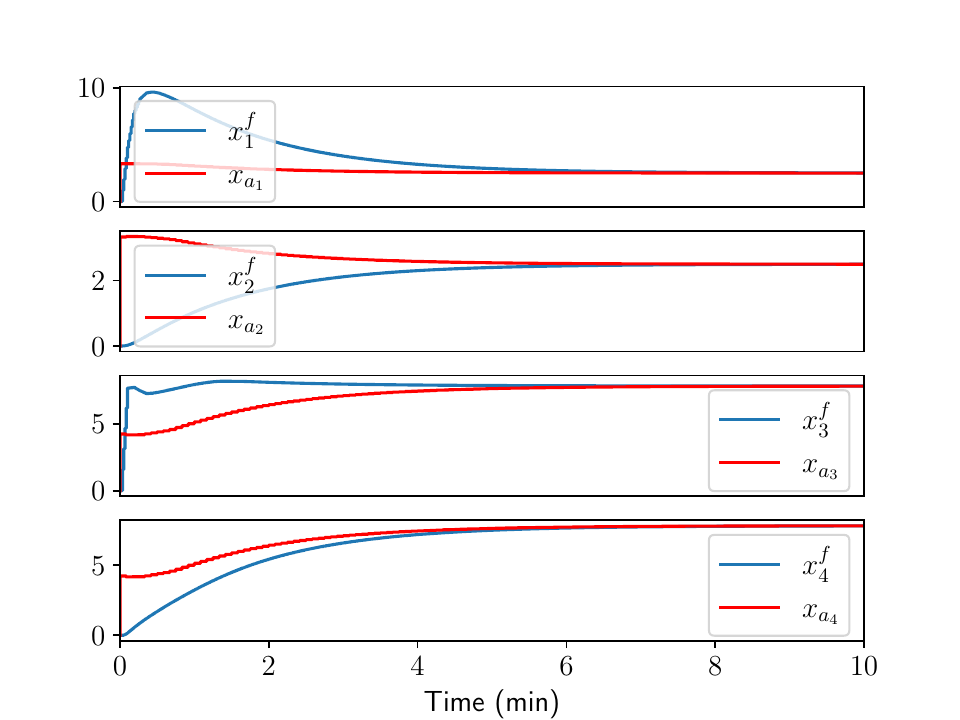}
         \vspace{-0.5cm}
         \caption{Fast states evolution.}
         \label{fig:fast_state}
      \end{figure}

      In Fig.~\ref{fig:bis} we can notice that indeed the output is steered to the desired reference $50$.
      Fig.~\ref{fig:input} shows the different input profiles, we can see the evolution of $u$ which is the drug rates injected to the patient, 
      $v$ corresponds to the input computed by the MPC 
      and $v_a$ the steady inputs computed by the MPC.
      As we can see, after some time (around 5 minutes) $v$ converges to $v_a$, while the applied $u$ has an offset due to the disturbance rejection term.
      Furthermore, Fig.~\ref{fig:fast_state} shows that the fast states converge to their respective steady-states.

\section{Conclusion}\label{Section:Conclusion}
   This paper proposed a new MPC formulation for anesthesia multiple input control,
   formulated as a linear MPC for tracking problem. 
   % Under a mild assumption on the output model, the control problem is formulated in a linear MPC for tracking setting.
   This formulation guarantees recursive feasibility and asymptotic stability and offers more flexibility by optimizing the steady state online.
   This opens new perspectives to apply techniques from economic MPC in order to guide the evolution of the equilibrium chosen by the MPC by 
   exploiting the additional cost term $V_d(\cdot)$. 
   To illustrate this, we can think of a scenario where one drug is more expensive than the other but allows a faster induction.
   Therefore, it can be useful to use it at first to join the target by reaching the equilibrium set as fast as possible. Once in the set, the MPC 
   could navigate in it to reach an equilibrium point where the expensive drug is spared.
   
   Moreover, in this paper the states are considered measured, but as in \cite{moussaDataBasedExtendedMoving2023} we can consider estimation approaches for future works, and apply techniques from robust or stochastic MPC
   that will still guarantee asymptotic stability and recursive feasibility.
   Further works should focus on extending this control law to take into account parametric uncertainties and noisy measurement to go closer
   to practically relevant research.

\begin{ack}
The authors warmly thank Benjamin Meyer and Rémi Wolf,
both anesthesiologists at Grenoble Hospital (France),
for their insightful discussions on closed-loop control strategies for anesthesia.
\end{ack}

\bibliography{ifacconf}             % bib file to produce the bibliography
                                                     % with bibtex (preferred)
%% There are a number of predefined theorem-like environments in
%% ifacconf.cls:
%%
%% \begin{thm} ... \end{thm}            % Theorem
%% \begin{lem} ... \end{lem}            % Lemma
%% \begin{claim} ... \end{claim}        % Claim
%% \begin{conj} ... \end{conj}          % Conjecture
%% \begin{cor} ... \end{cor}            % Corollary
%% \begin{fact} ... \end{fact}          % Fact
%% \begin{hypo} ... \end{hypo}          % Hypothesis
%% \begin{prop} ... \end{prop}          % Proposition
%% \begin{crit} ... \end{crit}          % Criterion
%% \begin{pf} ... \end{pf}              % proofs
\end{document}